# A deep learning approach to the texture optimization problem for friction control in lubricated contacts

Alexandre Silva[1,2], Veniero Lenzi[*,1,2], Sergey Pyrlin[1,2], Sandra Carvalho[3], Albano Cavaleiro[3], Luís Marques[1,2]

[1] *Physics Center of Universities of Minho and Porto (CF-UM-UP), University of Minho, Campus de Gualtar, 4710-057, Braga, Portugal*

[2] *Laboratory of Physics for Materials and Emergent Technologies, LapMET, University of Minho, 4710-057 Braga, Portugal*

[3] *University of Coimbra, CEMMPRE - Centre for Mechanical Engineering Materials and Processes, Department of Mechanical Engineering, Rua Luís Reis Santos, 3030-788, Coimbra, Portugal*

\* Corresponding author: Veniero Lenzi, E-mail: veniero.lenzi@fisica.uminho.pt

**Abstract:** The possibility to control friction through surface micro texturing could offer invaluable advantages in many fields, from wear and pollution reduction in the transportation industry to improved adhesion and grip. Unfortunately, the texture optimization problem is very hard to solve using traditional experimental and numerical methods, due to the complexity of the texture configuration space. In this work, we apply machine learning techniques to perform the texture optimization, by training a deep neural network to predict, with extremely high accuracy and speed, the Stribeck curve of a textured surface in lubricated contact. The deep neural network was used to completely resolve the mapping between textures and Stribeck curves, enabling a simple method to solve the texture optimization problem. This work demonstrates the potential of machine learning techniques in texture optimization for friction control in lubricated contacts.

# 1 Introduction

Our world is overwhelmed by the environmental impact of human activity and there is an imperative need to reduce pollution and mitigate its effects to avoid an irreversible global warming. The transportation industry, one of the largest contributors to polluting emissions, wastes a significant part of fuel and energy in overcoming friction forces between moving parts in contact [1], meaning that any solution to reduce friction would provide huge environmental and economic benefits. Because of this, research on friction reduction has always been at the forefront of tribology research and many possible solutions exist, such as the application of surface coatings [2] and the use of more performing and environmentally friendly lubricant formulations [3]. One of the most promising ways to control the friction between contacting surfaces is provided by surface texturing, a process that is increasingly more efficient due to significant processing advances [4] allowing for rapid generation of patterned surfaces. It is well known that a fine control of friction through surface texturing can be achieved in nature. For example, sharks are covered in a regular array of denticles which help to achieve drag reduction [5]. The same reduction has been seen in the skin of snakes and certain lizards that developed scales to reduce dry contact friction [6]. Specific nano-hierarchically structured patterns found in the feet of tree toads [7, 8] and geckos [9] have been shown to provide a strong boundary friction, granting them better grip on vertical surfaces. In engineering applications, many different kinds of nature-inspired patterns have also been tested for friction control [8]. However, the design of these textures is in general based on trial-and-error methods, meaning that the optimal texture for a specific application is extremely hard to find. From an experimental perspective, textured samples need to be fabricated and tested, thus optimizing a specific pattern would require an extensive sampling of the texture parameter space, resulting in time and resource costs that are prohibitive [10, 11]. The same problem occurs when using numerical approaches to evaluate the tribological performance of a system, where the Stribeck curve [12] is calculated by solving the Reynolds equation [13, 14], for multiple sliding speeds, coupled with a model for treating the contact friction [15]. Even if the simulation process is faster as a whole when compared to a single experiment, the calculations still require typically minutes to complete, meaning that our ability to sample the possible configuration space is incredibly limited [16]. Moreover, the relationship between patterns and resulting Stribeck curves is expected to be highly non-linear, based on current experimental and numerical understanding [4, 17, 18, 19]. A possible solution to the apparently insurmountable texture optimization problem might be offered by machine learning techniques. Machine learning (ML) encompasses a large range of algorithms and modeling tools used for large data processing tasks [20, 21] with typical applications being classification and regression problems in information technology [22, 23]. One of the most prominent ML techniques is represented by deep neural networks (DNN), which are used with considerable success in many fields of physics, from applications in condensed matter [24, 25] and materials science [26, 25] to the solution of complex nonlinear equations [27, 28]. Machine learning and artificial intelligence techniques, such as DNNs [29], have been recently introduced in tribology, where they found applications in many different areas [30]. What makes DNNs particularly appealing for the texture optimization problem is their universal approximation capability [22, 23], coupled with their extreme speed when compared to traditional methods [28]. As a matter of fact, it has been recently shown the classical Reynolds equation can be solved by means of a physics informed neural network [31]. In texture optimization problems, DNNs have been used to optimize the features of periodic patterns of nanopillars in optic metamaterials to achieve the desired properties, i.e., high electromagnetic wave absorption in some frequency windows [28]. In these works, a DNN replaced the Maxwell equations solver, and it could predict -in millisecond time- an absorbance spectrum based solely on the periodic pattern features.

In this work we developed an effective method for the optimization of surface texturing patterns for friction applications based on a deep neural network. The DNN was designed and trained to accurately predict the Stribeck curve of a dimple textured surface, thus replacing the standard Reynolds equation solver in the solution of the forward problem. Moreover, to solve the inverse problem, a fast search-based approach was implemented to predict a set of candidate surface parameters (dimple pattern and dimple radius) that yield a set of closely matching Stribeck curves. The performance and accuracy of the DNN and the inverse approach were validated by comparing with the solutions provided by a numerical solver of the Reynolds and contact friction model equations.

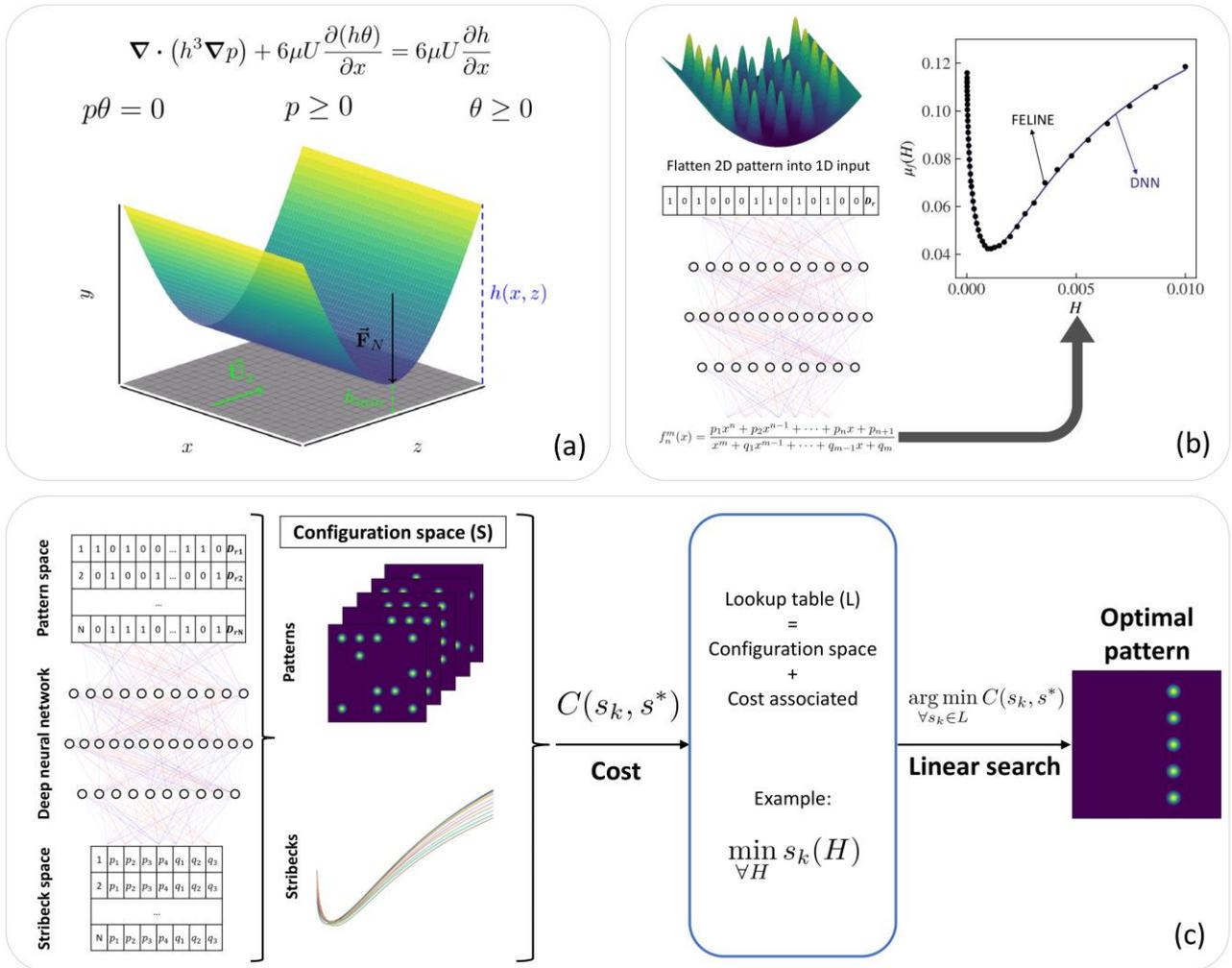

**Fig. 1** Schematic representation of the implementation of the DNN solution for the forward and reverse problem in texture optimization. (a) Non-conformal contact of surfaces modeled as a height profile function h(x, z), subject to load F moving relative to each other with velocity U. (b) Machine learning approach to predict the Stribeck curve of a textured surface, defined as the forward problem.

The two-dimensional texture is flattened into a 26-parameter input defining the arrangement of dimples and their radius, while the output is a set of 7 distinct parameters that allow the reconstruction of the Stribeck curve. (c) Machine learning approach to solve the texture optimization problem (inverse problem). The Stribeck curves of the full configuration space of pattern and dimple radius are obtained by using the forward DNN. Then, a lookup table is constructed after calculating an associated cost for each case, and a simple linear search is performed to find the optimal pattern according to the given cost function.

## 2 Methods

### 2.1 Solving the Reynolds equation

For any lubricated contact of height profile $h \equiv h(x, y)$ between two surfaces moving with relative speed $U$ subject to an external load $F$ and lubricant viscosity μ, as it is schematically represented in Fig. 1(a), we obtained the pressure profile p within the lubricant by solving the Reynolds equation, derived from the Navier-Stokes equations [14] with constant viscosity and temperature. To consider cavitation, that is, the possibility of formation of vapor filled cavities in the lubricant film [32, 33, 34, 35], we introduce a system of equations for the pressure p and cavitation fraction θ profiles:

$$\nabla \cdot (h^3 \nabla p) + 6\mu U \frac{\partial (h\theta)}{\partial x} = 6\mu U \frac{\partial h}{\partial x} \quad (1)$$

$$p\theta = 0 \quad (2)$$

$$p \geq 0 \quad (3)$$

$$\theta \geq 0 \quad (4)$$

The system of Eqs. (1)-(4) represents a linear complementarity problem (LCP) [36, 37, 38], which was solved using the inexact Newton (INE) method [39] by providing a restructuring of the system of equations into a damped Newton iteration. The INE method ensures that the solution follows the non-negativity conditions at every iteration, thus providing a correct physical description of cavitation boundaries [16].

Depending on sliding speed, applied load and viscosity, the system admits three different regimes of lubricated contact: boundary, mixed and hydrodynamic, which differ in their main friction mechanisms. In this work only the mixed and hydrodynamic regimes have been considered since they occur in the presence of lubricant within the contact whereas in the boundary regime the surfaces are in direct contact without lubricant in-between (dry friction). To treat the friction forces in the mixed and hydrodynamic regimes, we adopted the Greenwood-Tripp (GT) contact model [15, 40], which takes in account the roughness of the contacting surfaces. The total friction force is

$$f_t = f_{hydro} + f_{GT}. \quad (5)$$

The first term, which represents the hydrodynamic component of friction, depends on the pressure gradient generated within the lubricant film, written as

$$f_{hydro} = \left(\frac{h}{2}\frac{\partial p}{\partial x} + \frac{U\mu}{h}\right)A, \quad (6)$$

where $A$ is the total surface area of the lubricated contact. The second term of Eq. (5), which represents the contact component of friction, relies on experimentally fitted surface roughness parameters $\eta k \sigma$ defined in [15] to accurately predict friction. A table with the numerical values of the parameters used in data generation for the training of the deep neural network application is available in Section 3 of SI.

An open-source finite elements implementation of the solver for the system of Eqs. (1)-(6), FELINE [41], was specifically developed and used to generate the training data and validate DNN results. Owing to intrinsic limitations of the Reynolds equation, only non-conformal contacts with a parabolic shape, defined by its parabolic edge E0 are considered in this work. Further details on the contact geometry can be found in SI.

**2.2 Stribeck curve calculation**

To obtain the Stribeck curve, one must compute the coefficient of friction (COF) for each relevant sliding speed as an integral of the total friction force in (5), written as:

$$COF(H) = \frac{1}{FA} \int_\Omega f_t(H) \, d\Omega, \qquad (7)$$

where

$$H = \frac{\mu U}{F} \qquad (8)$$

is a dimensionless parameter dependent of the relative sliding speed termed Hersey number.

Importantly, these computations are only performed when the friction generated lift, which is a function of the minimum distance between the surfaces, and the applied load are in balance for each Hersey number. A total of 50 different Hersey number values, equally spaced in a logarithmic scale, were used for each Stribeck curve in the range $H \in [10^{-5}, 10^{-2}]$.

**2.3 Design and training of the DNN**

The textured surface in a lubricated contact is defined by a set of parameters: the dimple map $D_{map}$ of dimension $(D_x, D_y)$ which encodes the placement of dimples on the surface (see Section 1 of S.I.), the dimple depth $D_d$, the dimple radius $D_r$, the parabolical edge $E_0$, and the surface roughness parameters $\eta k \sigma$. After numerical evaluation (see Section 2 of S.I.), the value of dimple depth was fixed to $D_d^0 = 6 \, \mu m$ because its impact on the Stribeck curves is much less pronounced than that of the dimple radius variation. Hence, for the case study presented here only the dimple radius and texture pattern were allowed to change, with all the dimples in the pattern having the same depth $D_d^0$.

We considered a $5 \times 5$ grid of dimples with 6 possible $D_r$ in the interval $[40, 60] \, \mu m$, thereby consisting in 26 parameters as our network input. This $D_r$ interval was selected since it provides a sufficient range for optimization while remaining within the validity conditions of the Reynolds equation. Even if such a configuration space appears simple at a first glance, it contains a total of $N = 6 \times 2^{25} \approx 2.01 \times 10^8$ possible texture configurations, rendering the texture optimization problem impossible to solve for any traditional solution approach.

In order to represent all Stribeck curves in the configuration space with the same number of parameters, a rational fit of the curves calculated with FELINE, defined as

$$f_n^m(x) = \frac{p_1 x^n + p_2 x^{n-1} + \cdots + p_n x + p_{n-1}}{x^m + q_1 x^{m-1} + \cdots + q_{m-1} x + q_m}, \quad (9)$$

with polynomial degrees $(n, m) = (3, 3)$ was found to be the best compromise between accuracy and total number of parameters when representing a Stribeck curve. As a result, the DNN output consists of just 7 parameters.

Regarding the DNN architecture, we adopted a simple topology consisting of 6 fully connected hidden layers with a number of neurons {32, 64, 96, 96, 64, 32} under no regularization using the ReLU activation function with He normal initialization [42] due to its performance and simplicity [43, 44]. Also, since the number of hidden layers in our network is relatively small, we do not expect the notorious "dying ReLU problem" in this study [45]. The optimizer of choice was Nadam as it incorporates Nesterov momentum which can improve the convergence of the learning process [46, 47, 48]. The RMSE of the predicted rational fit coefficients was used as the network loss function. The training and testing process was completely done using the Keras high-level API [49] of TensorFlow version 2 [50].

The DNN training set was populated by randomly sampling sets of dimple maps $\boldsymbol{D_{map}}$, whose corresponding patterns were solved for all the $D_r$ values to obtain the corresponding Stribeck curves. In total, around 60000 different combinations of patterns and dimple radii were computed using the FELINE solver, which required 6 days of computation time on 300 simultaneously running processes on Intel(R) Xeon(R) CPU E5-2697 v2 cores.

Owing to the fact that the boundary pressure is the same at $y = 0$ and $y = L$, we expect that a mirror reflection of any pattern over the $y = L/2$ axis does not change the corresponding Stribeck curve. This symmetry was explicitly included in the dataset by assigning the same Stribeck curve both to a pattern and to its reflection. This step is important in enhancing the overall physical accuracy of the DNN, while requiring no additional generation of data. From the generated dataset we selected 10% as a validation set, thus our resulting training set contains 54000 pairs of surface parameters and Stribeck curves and accounts for only 0.05% of the total configuration space.

## 3 Results and discussion

### 3.1 Solution of the forward problem

The forward problem, schematically represented in Fig. 1(b), was solved and examples of the DNN predictions are shown in Fig. 2 for a few cases in the validation set compared to data produced with the FELINE solver. The median RMSE of cases in the validation set is $5.7 \times 10^{-4}$, meaning that the network predictions are very accurate and show no appreciable difference with the Stribeck curves calculated with FELINE.

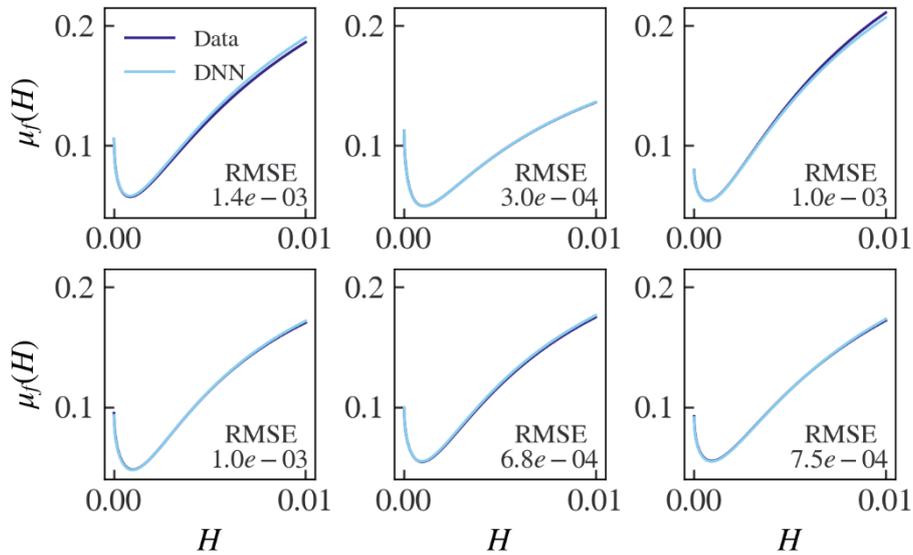

**Fig. 2** Network prediction of the Stribeck curves for randomly selected patterns in the validation set. The prediction accuracy is evaluated in terms of the root mean squared error (RMSE) in comparison to true data.

Fig. 3(a) shows a histogram of the RMSE distribution for the validation set predictions of the full Stribeck curve and, for two separate regimes of the Stribeck curve, that is the mixed regime and the hydrodynamic regime. To correctly establish boundaries for these regimes we used the lambda parameter criteria [51]:

$$\lambda = \frac{h_{min}}{\sigma}, \tag{10}$$

where $h_{min}$ denotes the minimum thickness of the lubricant film (or minimum distance between the contacting films) and $\sigma$ is one of the surface roughness parameters. For $\lambda > 3$ the contact regime is said to be hydrodynamic, while the mixed regime occurs for $1 < \lambda < 3$. After taking an average of $\lambda$ for all curves in the validation set we found that the averaged value $H = 0.0015$ represents well the point in which the lubrication regime changes. Therefore, for $H \in [0, 0.0015]$ we have the mixed regime and for $H \in [0.0015, 0.01]$ we have the hydrodynamic regime.

The corresponding median, 95th and 99th percentile of the different histograms is reported in table 1. Low RMSE values ($< 10^{-3}$) are consistently encountered in all regimes, indicating that the trained DNN is reliable across all the data. However, a better accuracy of the DNN in the mixed region was observed, compared to the hydrodynamic region. This is likely due to the larger span of COF values in the hydrodynamic region, for the same number of training samples, resulting in a lower accuracy prediction of the DNN therein.

**Table 1** Median, 95th and 99th percentile of the RMSE values shown in the histogram in Fig. 3(b).

| Regime | Median | 99% | 95% |
| --- | --- | --- | --- |
| Mixed | $2.4 \times 10^{-4}$ | $7.9 \times 10^{-4}$ | $6.0 \times 10^{-4}$ |
| Hydrodynamic | $9.4 \times 10^{-4}$ | $6.1 \times 10^{-3}$ | $3.8 \times 10^{-3}$ |
| Total | $5.7 \times 10^{-4}$ | $3.4 \times 10^{-3}$ | $2.1 \times 10^{-3}$ |

To assess the quality of the trained network, it is also important to verify its ability to interpolate and extrapolate results in terms of the dimple radius, since it was trained only with 6 possibilities for it. In this regard, we compared both the network and the FELINE solver solutions for values in-between and outside the dimple radius interval 40 μm to 60 μm.

A pattern was randomly picked and its corresponding Stribeck curve was computed with the DNN and FELINE in order to obtain a RMSE of their difference, which was plotted in Fig. 3(c) as a function of $D_r$. As one can see in Fig. 3(c), in region (II), the interpolation region, there is a small difference between the interpolation results from the DNN and the corresponding ones from the FELINE solver, showing that the network is capable of accurate interpolating behavior.

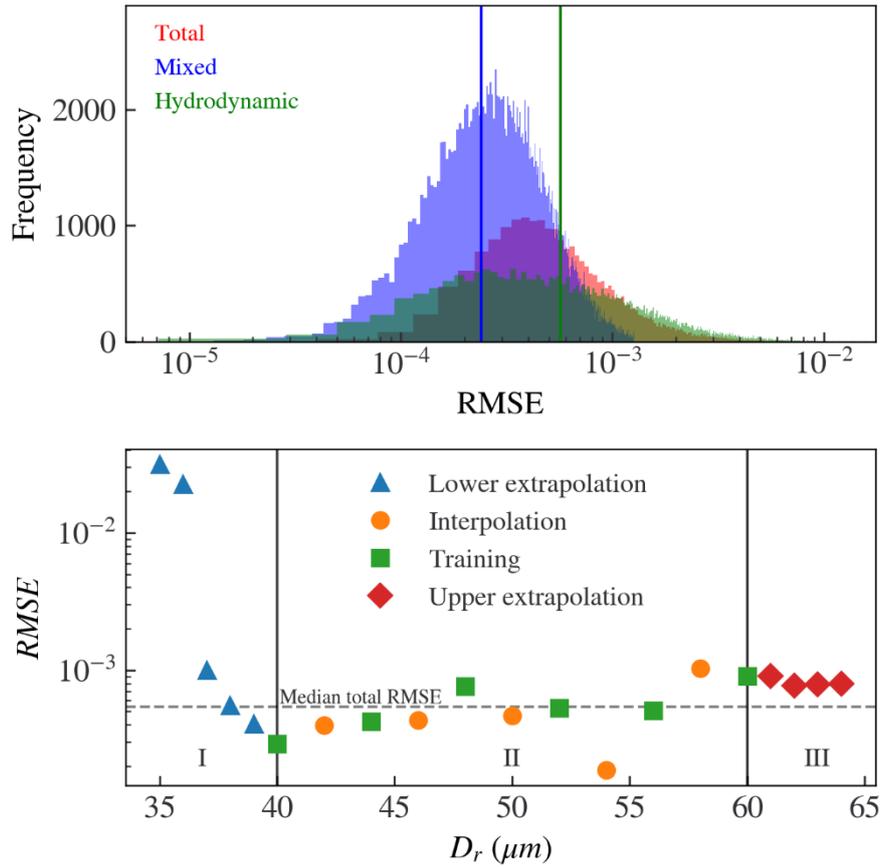

**Fig. 3** (a) Histogram of the RMSE for all patterns in the validation set in different regimes. The medians for the mixed, hydrodynamic, and total regions are also shown. (b) Interpolation and extrapolation study of dimple radius versus RMSE where: (I) indicates the lower extrapolation bound $D_r \in [35, 39]$, (II) the interpolated values $D_r \in [42, 46, 50, 54, 58]$ and the values used in training, (III) the upper extrapolation bound $D_r \in [61, 65]$.

For the extrapolation cases, regions (II) and (III), we see that the lower extrapolation bound works significantly worse than the upper extrapolation bound. In the hydrodynamic regime of lubrication, the COF increases linearly with increasing Hersey number. Contrary to this, in the mixed regime of lubrication, the COF increases exponentially with decreasing Hersey number. Extrapolation is typically more accurate for linear behavior, hence resulting in a larger extrapolation error for the lower bound of extrapolation.

In terms of timing, the DNN is $10^6$ times faster when compared to the FELINE solver. In conclusion, we have successfully designed a DNN that meets the requirements of speed and accuracy needed to fully solve the texture optimization problem for tribological applications.

**3.2 Solution of the inverse problem**

By using the DNN developed to solve the forward problem it is possible to generate a list of Stribeck curves for all the possible textures within the configuration space $S$. A lookup table $L$ is then constructed by applying to each case a cost function $C(s_k, s^*)$ where $s_k$ is a particular Stribeck curve in the configuration space and $s^*$ relates to some aspects of the Stribeck curve that will be searched for. This lookup table can then be used to efficiently search for the texture that yields a Stribeck curve with the desired features. This process is summarized in Fig. 1(c).

As a first example of the power of this method, let us consider the following cost function to generate $L$,

$$C(s_k, s^*) = \min_{\forall H} s_k(H), \tag{11}$$

which simply finds the minimum COF of a particular Stribeck curve $s_k \in S$. It is then possible to find the smallest COF of every Stribeck curve in the configuration space, and thus the pattern that generates it, by performing the following linear search through the lookup table:

$$\{p, s\}^{min} = \arg\min_{\forall s_k \in L} C(s_k, s^*). \tag{12}$$

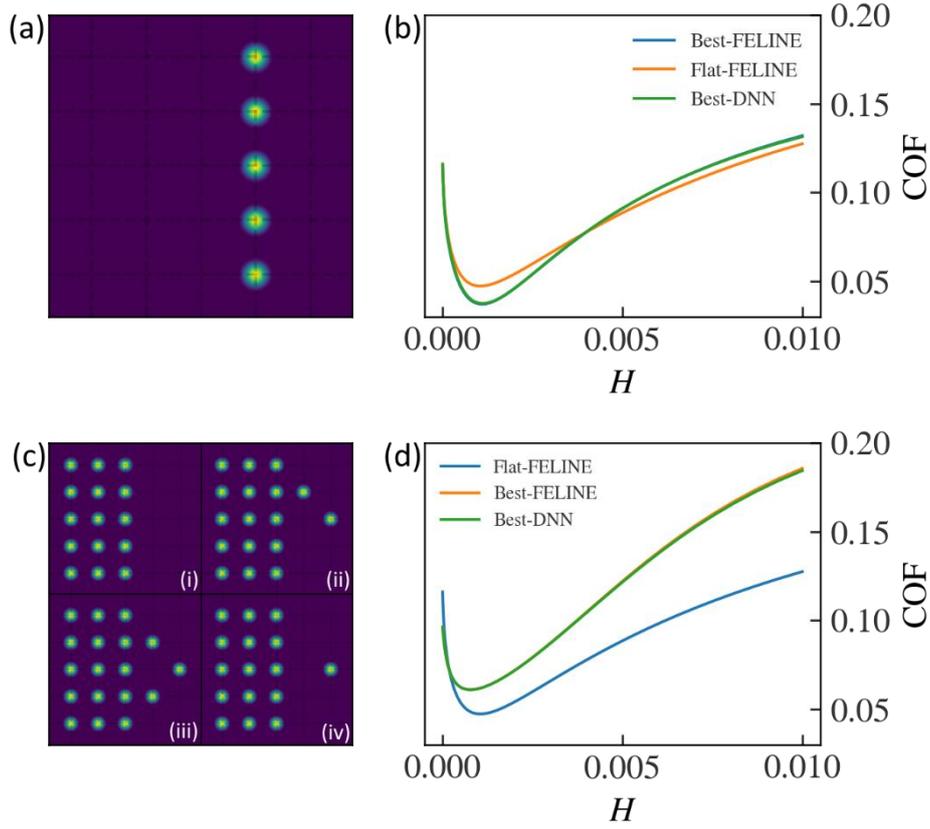

**Fig. 4** Results of the linear search approach on the solved full configuration space of textures used to solve the texture optimization problem. (a) Optimal pattern that yields the Stribeck curve with the smallest minimum, according to the cost function search in Eq. (12). The corresponding Stribeck curve, calculated with both FELINE and the DNN, is reported in panel (b), where it is compared with the untextured case. (c) Textures that yield nearly matching Stribeck curves with the largest minimum, according to the cost function search in Eq. (13). The corresponding Stribeck curve, calculated with both FELINE and the DNN, is reported in panel (d), where it is compared with the untextured case.

The resulting optimal pattern obtained using this method is shown in Fig. 4(a), additionally the corresponding Stribeck curve (solved with FELINE and the DNN) is reported in Fig. 4(b) in comparison to the untextured case. Importantly, the results obtained with FELINE and the DNN are in great agreement. The position of the dimples, right after contact region at $x = 0.5L$, provides a reduction in the size of the cavitation region in the mixed regime, lowering the overall friction of the system in this regime.

Similarly, one can perform the linear search that finds the Stribeck curve with the highest minimum in the lookup table, more formally

$$\{p, s\}^{max} = \arg\max_{\forall s_k \in L} C(s_k, s^*). \qquad (13)$$

A collection of similar patterns is shown in Fig. 4(c). Since the corresponding Stribeck curves (solved with FELINE and the DNN) for each of the patterns (i)-(iv) are nearly identical, only the Stribeck curve of pattern (i) is reported in Fig. 4(d) in comparison to the untextured case. Most importantly, a trend was observed for the patterns which maximized the value of the minimum, being the placement of 3 columns of dimples before the contact region. The near identical nature of the Stribeck curve results for patterns (i)-(iv) highlights an important feature of this method, that is the potential of finding multiple solutions to the same problem. This can be extremely important from an experimental point of view because, at parity of performance, a particular texture may be better in terms of processing in a laboratory.

By considering a different cost function, one can search for the optimal texture according to more general aspects of the Stribeck curve, such as the minimization of friction in the hydrodynamic range. To do this, we consider the following cost function

$$C(s_k, s^*) = \int_{H_1}^{H_2} [s_k(H) - s_0(H)] \, dH, \qquad (14)$$

where $H_1 = 0.0015$ and $H_2 = 0.01$, encompassing the hydrodynamic range. This cost function computes the area under the Stribeck curve in the hydrodynamic region for some Stribeck curve $s_k \in L$ minus the area under the Stribeck curve of the untextured surface, $s_0$.

By performing the same linear search as in (12), one can determine the best result for this problem, which is represented in Fig. 5(a). The search for the smallest overall COF in the hydrodynamic range has additionally found a pattern that reduces friction in every region of the Stribeck curve.

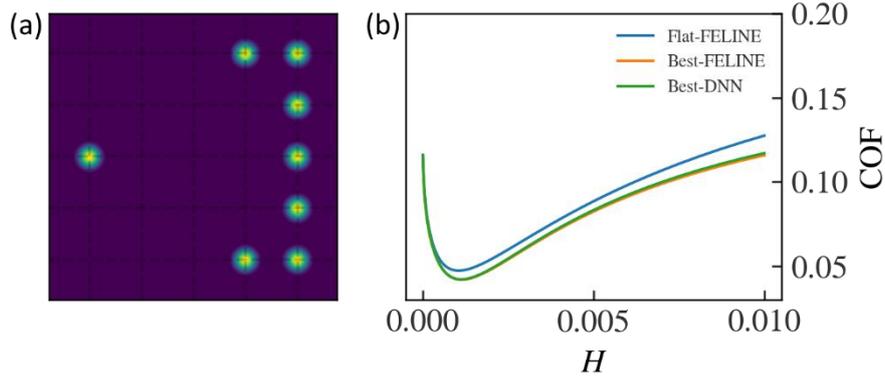

**Fig. 5** Results of the linear search approach on the solved full configuration space of textures used to solve the texture optimization problem. (a) Optimal pattern that yields the Stribeck curve with the overall smallest COF in the hydrodynamic regime. The corresponding Stribeck curve, calculated with both FELINE and the DNN is reported in panel (b), where it is compared with the untextured case.

The above examples demonstrate how the texture optimization problem turns in to a very simple task using the DNN, which allowed for a very efficient search of the optimal pattern in the full configuration space, with the desired friction characteristics.

## 4 Conclusions

We have successfully designed and trained a deep neural network capable of accurately predicting the resulting Stribeck curve generated by a dimpled texture with median root mean square errors of $5.7 \times 10^{-4}$. This type of texture, composed of an array of $5 \times 5$ possible dimples with dimple radius $D_r$ has an unpredictable and highly non-linear effect on the surface friction coefficient. The DNN can efficiently compute all possible cases of a total of around 100 million possibilities, trained with only 0.05% of them, thus enabling us to solve the texture optimization problem which is otherwise impossible to treat by traditional experimental and numerical methods. We determined both extremes of an optimization problem by taking advantage of the incredible performance of our DNN, predicting the relevant optimal textures in the process. We investigated properties of the developed DNN such as accuracy, extrapolation, and interpolation capabilities, demonstrating its robustness and reliability. This work paves the way for the use of deep learning as a tool to realize careful friction control of surfaces through optimally designed textures.

**Acknowledgements**

This work was supported by the Portuguese Foundation for Science and Technology (FCT, I.P.) in the framework of the Strategic Funding UIDB/04650/2020, projects PTDC/EME-SIS/30446/2017, project POCI-01–0247-FEDER-045940, and Advanced Computing Project CPCA/A2/4513/2020 for accessing MACC-BOB HPC resources.

**Declaration of competing interest**

The authors have no competing interests to declare that are relevant to the content of this article.

**Electronic Supplementary Material (ESM)**

Electronic Supplementary Material: Supplementary material (geometry of the lubricated contact, effect of dimple radius and dimple depth on Stribeck curves, parameters used in data generation for the deep neural network training set) is available in the online version of this article.

**References**

[1] K. Holmberg, P. Andersson, A. Erdemir, Global energy consumption due to friction in passenger cars, Tribology International 47 (2012) 221–234. doi:https://doi.org/10.1016/j.triboint.2011.11.022.

[2] N. Shaigan, W. Qu, D. G. Ivey, W. Chen, A review of recent progress in coatings, surface modifications and alloy developments for solid oxide fuel cell ferritic stainless steel interconnects, Journal of Power Sources 195 (6) (2010) 1529–1542. doi:https://doi.org/10.1016/j.jpowsour.2009.09.069.

[3] A. E. Somers, P. C. Howlett, D. R. MacFarlane, M. Forsyth, A review of ionic liquid lubricants, Lubricants 1 (1) (2013) 3–21. doi:10.3390/lubricants1010003.

[4] I. Etsion, State of the Art in Laser Surface Texturing , Journal of Tribology 127 (1) (2005) 248–253. doi:10.1115/1.1828070.

[5] G. Liu, Z. Yuan, Z. Qiu, S. Feng, Y. Xie, D. Leng, X. Tian, A brief review of bio-inspired surface technology and application toward underwater drag reduction, Ocean Engineering 199 (2020) 106962. doi:https://doi.org/10.1016/j.oceaneng.2020.106962.

[6] C. Greiner, M. Schäfer, Bio-inspired scale-like surface textures and their tribological properties, Bioinspiration amp Biomimetics 10 (2015) 044001. doi:10.1088/1748-3190/10/4/044001.

[7] T. Endlein, A. Ji, D. Samuel, N. Yao, Z. Wang, W. J. P. Barnes, W. Federle, M. Kappl, Z. Dai, Sticking like sticky tape: tree frogs use friction forces to enhance attachment on overhanging surfaces, Journal of the Royal Society, Interface 10 (80) (Jan 2013). doi:10.1098/rsif.2012.0838.

[8] L. Zhang, H. Chen, Y. Guo, Y. Wang, Y. Jiang, D. Zhang, L. Ma, J. Luo, L. Jiang, Micro–nano hierarchical structure enhanced strong wet friction surface inspired by tree frogs, Advanced Science 7 (20) (2020) 2001125. doi:https://doi.org/10.1002/advs.202001125.

[9] A. Y. Stark, I. Badge, N. A. Wucinich, T. W. Sullivan, P. H. Niewiarowski, A. Dhinojwala, Surface wettability plays a significant role in gecko adhesion underwater, Proceedings of the National Academy of Sciences of the United States of America 110 (16) (Apr 2013). doi:10.1073/pnas.1219317110.

[10] M. Marian, A. Almqvist, A. Rosenkranz, M. Fillon, Numerical microtexture optimization for lubricated contacts—a critical discussion, Friction (Apr 2022). doi:10.1007/s40544-022-0609-6.
URL https://doi.org/10.1007/s40544-022-0609-6

[11] N. Myshkin, M. Petrokovets, A. Kovalev, Tribology of polymers: Adhesion, friction, wear, and mass-transfer, Tribology International 38 (11-12) (2005) 910–921.

[12] R. Stribeck, Characteristics of plain and roller bearings., Zeit. VDI, 46 (1902).

[13] A. Akchurin, R. Bosman, P. M. Lugt, M. van Drogen, On a model for the prediction of the friction coefficient in mixed lubrication based on a loadsharing concept with measured surface roughness, Tribology Letters 59 (1) (2015) 19.

[14] O. Reynolds, Iv. on the theory of lubrication and its application to mr. beauchamp tower's experiments, including an experimental determination of the viscosity of olive oil, Philosophical transactions of the Royal Society of London (177) (1886) 157–234.

[15] J. A. Greenwood, J. H. Tripp, The contact of two nominally flat rough surfaces, Proceedings of the Institution of Mechanical Engineers 185 (1) (1970) 625–633. doi:10.1243/PIME_PROC_1970_185_069_02.

[16] D. Gropper, L. Wang, T. J. Harvey, Hydrodynamic lubrication of textured surfaces: A review of modeling techniques and key findings, Tribology International 94 (2016) 509–529. doi:https://doi.org/10.1016/j.triboint.2015.10.009.

[17] C. Greiner, T. Merz, D. Braun, A. Codrignani, F. Magagnato, Optimum dimple diameter for friction reduction with laser surface texturing: The effect of velocity gradient, Surface Topography: Metrology and Properties 3 (2015) 044001. doi:10.1088/2051-672X/3/4/044001.

[18] R. Ausas, P. Ragot, J. Leiva, M. Jai, G. Bayada, G. C. Buscaglia, The Impact of the Cavitation Model in the Analysis of Microtextured Lubricated Journal Bearings, Journal of Tribology 129 (4) (2007) 868–875. doi:10.1115/1.2768088.

[19] R. F. Ausas, M. Jai, G. C. Buscaglia, A mass-conserving algorithm for dynamical lubrication problems with cavitation, Journal of Tribology 131 (3) (jun 2009). doi:10.1115/1.3142903.

[20] G. Carleo, I. Cirac, K. Cranmer, L. Daudet, M. Schuld, N. Tishby, L. VogtMaranto, L. Zdeborová, Machine learning and the physical sciences, Rev. Mod. Phys. 91 (2019) 045002. doi:10.1103/RevModPhys.91.045002.

[21] M. Biehl, A. Mietzner, Statistical mechanics of unsupervised learning, Europhysics Letters (EPL) 24 (5) (1993) 421–426. doi:10.1209/02955075/24/5/017.

[22] K. Hornik, M. Stinchcombe, H. White, Multilayer feedforward networks are universal approximators, Neural Networks 2 (5) (1989) 359–366. doi:https://doi.org/10.1016/0893-6080(89)90020-8.

[23] G. Cybenko, Approximation by superpositions of a sigmoidal function, Mathematics of Control, Signals and Systems 2 (4) (1989) 303–314. doi:10.1007/BF02551274.

[24] J. Carrasquilla, R. G. Melko, Machine learning phases of matter, Nature Physics 13 (5) (2017) 431–434. doi:10.1038/nphys4035.

[25] A. J. Lew, C.-H. Yu, Y.-C. Hsu, M. J. Buehler, Deep learning model to predict fracture mechanisms of graphene, npj 2D Materials and Applications 5 (1) (2021) 48. doi:10.1038/s41699-021-00228-x.

[26] K. Guo, Z. Yang, C.-H. Yu, M. J. Buehler, Artificial intelligence and machine learning in design of mechanical materials, Mater. Horiz. 8 (2021) 1153–1172. doi:10.1039/D0MH01451F.

[27] J. Pathak, B. Hunt, M. Girvan, Z. Lu, E. Ott, Model-free prediction of large spatiotemporally chaotic systems from data: A reservoir computing approach, Phys. Rev. Lett. 120 (2018) 024102.
doi:10.1103/PhysRevLett.120.024102.

[28] C. C. Nadell, B. Huang, J. M. Malof, W. J. Padilla, Deep learning for accelerated all-dielectric metasurface design, Opt. Express 27 (20) (2019) 27523–27535.

[29] I. Argatov, Artificial neural networks (anns) as a novel modeling technique in tribology, Frontiers in Mechanical Engineering 5 (2019). doi:10.3389/fmech.2019.00030.
URL https://www.frontiersin.org/article/10.3389/fmech.2019.00030

[30] M. Marian, S. Tremmel, Current trends and applications of machine learning in tribology—a review, Lubricants 9 (9) (2021).
doi:10.3390/lubricants9090086.

[31] A. Almqvist, Fundamentals of physics-informed neural networks applied to solve the reynolds boundary value problem, Lubricants 9 (8) (2021). doi:10.3390/lubricants9080082.

[32] H. Elrod, A computer program for cavitation and starvation problems, Journal of Lubrication Technology (1975). doi:10.1115/1.3251669.

[33] B. Jakobsson, L. Floberg, The finite journal bearing, considering vaporization (das gleitlager von endlicher breite mit verdampfung), 1957.

[34] A. Almqvist, P. Wall, Modelling cavitation in (elasto)hydrodynamic lubrication, in: P. H. Darji (Ed.), Advances in Tribology, IntechOpen, Rijeka,
2016, Ch. 9. doi:10.5772/63533.

[35] M. Giacopini, M. T. Fowell, D. Dini, A. Strozzi, A Mass-Conserving Complementarity Formulation to Study Lubricant Films in the Presence of
Cavitation, Journal of Tribology 132 (4) (09 2010). doi:10.1115/1.4002215.

[36] R. W. Cottle, J.-S. Pang, R. E. Stone, The linear complementarity problem, SIAM, 2009.

[37] A. Almqvist, J. Fabricius, R. Larsson, P. Wall, A New Approach for Studying Cavitation in Lubrication, Journal of Tribology 136 (1) (11 2013). doi:10.1115/1.4025875.

[38] L. Bertocchi, D. Dini, M. Giacopini, M. T. Fowell, A. Baldini, Fluid film lubrication in the presence of cavitation: a mass-
conserving two-dimensional formulation for compressible, piezoviscous and non-newtonian fluids, Tribology International 67 (2013) 61–71.
doi:https://doi.org/10.1016/j.triboint.2013.05.018.

[39] F. Mezzadri, E. Galligani, An inexact newton method for solving complementarity problems in hydrodynamic lubrication, Calcolo 55 (1) (feb 2018). doi:10.1007/s10092-018-0244-9.

[40] R. Gohar, H. Rahnejat, Fundamentals of Tribology, IMPERIAL COLLEGE PRESS, 2008. doi:10.1142/p553.

[41] A. Silva, V. Lenzi, A. Cavaleiro, S. Carvalho, L. Marques, Feline: Finite element solver for hydrodynamic lubrication problems using the inexact newton method, Computer Physics Communications (2022)
108440. doi:https://doi.org/10.1016/j.cpc.2022.108440.

[42] K. He, X. Zhang, S. Ren, J. Sun, Delving deep into rectifiers: Surpassing human-level performance on imagenet classification, in: Proceedings of the IEEE international conference on computer vision, 2015, pp. 1026–1034.

[43] M. Capra, B. Bussolino, A. Marchisio, G. Masera, M. Martina, M. Shafique, Hardware and software optimizations for accelerating deep neural networks: Survey of current trends, challenges, and the road ahead, IEEE Access 8 (2020) 225134–225180. doi:10.1109/ACCESS.2020.3039858.

[44] A. F. Agarap, Deep learning using rectified linear units (relu), CoRR abs/1803.08375 (2018). arXiv:1803.08375.

[45] L. Lu, Y. Shin, Y. Su, G. Em Karniadakis, Dying relu and initialization: Theory and numerical examples, Communications in Computational Physics 28 (5) (2020) 1671–1706. doi:https://doi.org/10.4208/cicp.OA-20200165.

[46] S. J. Reddi, S. Kale, S. Kumar, On the convergence of adam and beyond, in: International Conference on Learning Representations, 2018.

[47] S. Salman, X. Liu, Overfitting mechanism and avoidance in deep neural networks (2019). arXiv:1901.06566.

[48] H. Tan, K. H. Lim, Vanishing gradient mitigation with deep learning neural network optimization, 2019, pp. 1–4. doi:10.1109/ICSCC.2019.8843652.

[49] F. Chollet, et al., Keras (2015).

[50] M. Abadi, A. Agarwal, P. Barham, E. Brevdo, Z. Chen, C. Citro, G. S. Corrado, A. Davis, J. Dean, M. Devin, S. Ghemawat, I. Goodfellow, A. Harp, G. Irving, M. Isard, Y. Jia, R. Jozefowicz, L. Kaiser, M. Kudlur, J. Levenberg, D. Mané, R. Monga, S. Moore, D. Murray, C. Olah, M. Schuster,

J. Shlens, B. Steiner, I. Sutskever, K. Talwar, P. Tucker, V. Vanhoucke,

V. Vasudevan, F. Viégas, O. Vinyals, P. Warden, M. Wattenberg, M. Wicke, Y. Yu, X. Zheng,

TensorFlow: Large-scale machine learning on heterogeneous systems, software available from tensorflow.org (2015).

[51] J. Hansen, M. Björling, R. Larsson, Lubricant film formation in rough surface non-conformal conjunctions subjected to gpa pressures and high slideto-roll ratios, Scientific Reports 10 (1) (2020) 22250. doi:10.1038/s41598-
020-77434-y.

**Author biography**

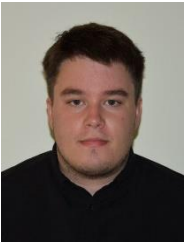

**Alexandre Silva.** He received his bachelor and MSc degree from University of Minho, Portugal in 2019 and 2022 respectively. Since 2022 he spent his time in the Laboratory of Physics for Materials and Emergent Technologies at University of Minho. His main scientific interest started with the simulation of electric discharges, during his MSc dissertation work he researched potential applications of machine learning for the optimization of tribological systems, and recently his focus is in DFT simulation in search of emergent properties in view for utilization in novel applications.

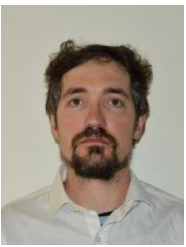

**Veniero Lenzi.** Veniero Lenzi is a postdoctoral researcher at the computational physics group at the physics centre of university of Minho. His area of research involves the application of ab-initio and classical molecular dynamics techniques to the study of condensed matter systems, including those of tribological interest such as solid lubricants in superhard nanocomposite matrix. Moreover, he recently worked on the application of finite element method techniques to study friction between texturized surfaces in lubricated contact.

**Graphical abstract**

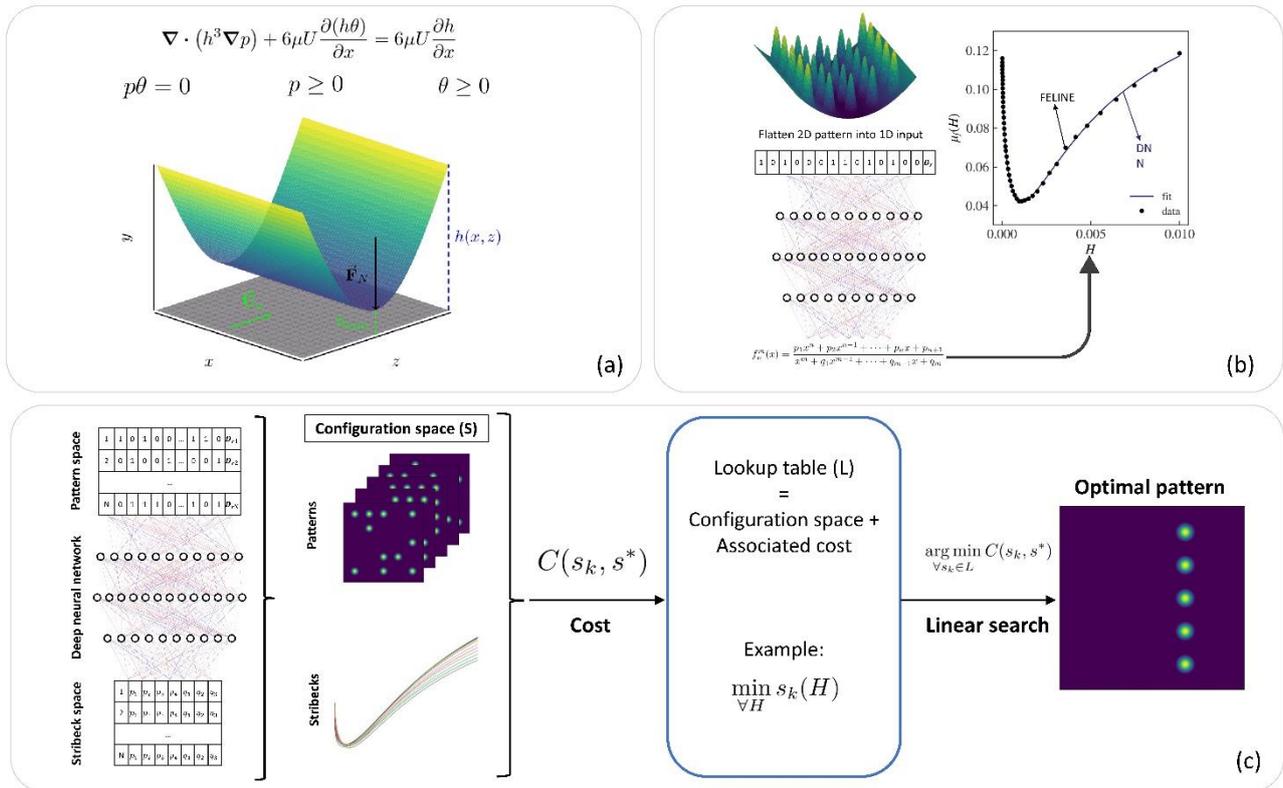

# Supplementary Information

## A deep learning approach to the texture optimization problem for friction control in lubricated contacts

Alexandre Silva[1,2], Veniero Lenzi[*,1,2], Sergey Pyrlin[1,2], Sandra Carvalho[3], Albano Cavaleiro[3], Luís Marques[1,2]

[1] *Physics Center of Universities of Minho and Porto (CF-UM-UP), University of Minho, Campus de Gualtar, 4710-057, Braga, Portugal*
[2] *Laboratory of Physics for Materials and Emergent Technologies, LapMET, University of Minho, 4710-057 Braga, Portugal*
[3] *University of Coimbra, CEMMPRE - Centre for Mechanical Engineering Materials and Processes, Department of Mechanical Engineering, Rua Luís Reis Santos, 3030-788, Coimbra, Portugal*

* Corresponding author: Veniero Lenzi, E-mail: veniero.lenzi@fisica.uminho.pt

## Section 1 – Geometry of the lubricated contact

Due to the intrinsic limitations of the Reynolds equation, the lubricated contact must be non-conformal (as shown in Figure S1), meaning that for a profile $h$ we must ensure thate

$$\frac{\partial h}{\partial x} \neq 0.$$

Hence, we assume that for any texture the contact shape is a parabolic contact $h_p(x, y)$ of the form

$$h_p(x, y) = E_0 \left(x - \frac{L_x}{2}\right)^2,$$

where $E_0$ represents the relative edge height of the profile. Additionally, the roughness parameters $\eta k \sigma$ obtained from the Greenwood-Tripp model, model the roughness of both surfaces.

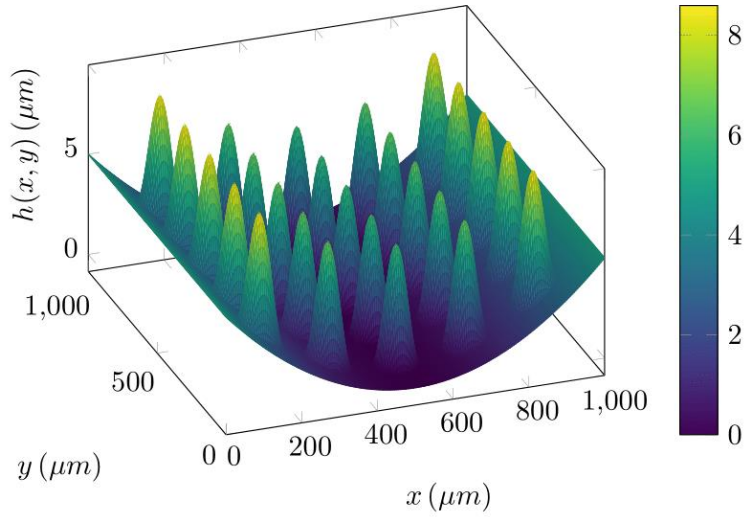

*Figure S1* Non-conformal height profile for a fully dimpled texture with dimple map dimensions $5 \times 5$.

To obtain the total height profile $h$ the part corresponding to the dimples needs to be added to $h_p(x, y)$. Consider a dimpled texture represented as a two-dimensional height profile of dimension $(L_x, L_y)$, where a $(D_x, D_y)$ array of regularly spaced dimples along $x$ and $y$ is placed. For any discrete pair of $(D_x, D_y)$ there exists a set of discrete coordinates $(x_i, y_j)$ for the dimple centers:

$$x_i = \frac{L_x}{2D_x} + i\frac{L_x}{D_x}$$

$$y_j = \frac{L_y}{2D_y} + j\frac{L_y}{D_y}$$

The dimple array admits a binary representation, thae dimple map $D_{map}$, which consists of a two-dimensional array of bits with $D_x \times D_y$ entries, where each entry specifies the existence - or not - of a dimple at coordinates $C_{ij} \equiv (x_i, y_j)$.

## Section 2 – Effect of dimple radius and dimple depth on Stribeck curves.

A fully dimpled texture, represented in figure S1, was studied for the effect of varying $D_r$ and $D_d$. We fixed $D_d = 6\ \mu m$ and allowed $D_r$ to change. We represent in figure S2 a set of Stribeck curves calculated using FELINE for different dimple radii within a Hersey number range $H \in [10^{-5}, 10^{-2}]$.

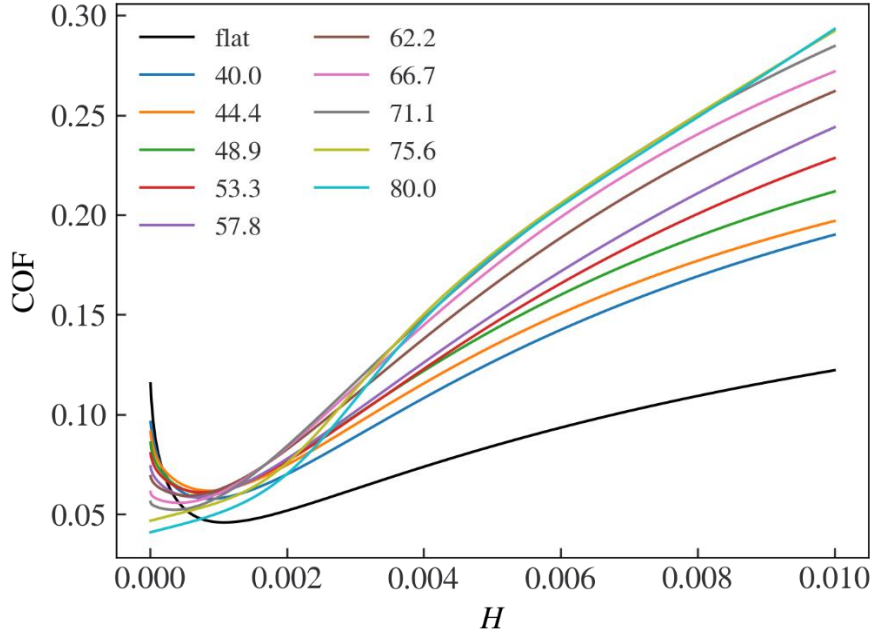

*Figure S2* Stribeck curves of the fully dimpled case with fixed dimple depth and changing dimple radius, along with the untextured case.

Taking the untextured case as reference, it is immediately obvious that the dimple radius has effects both in the mixed and hydrodynamic regimes of lubrication, up to 300% for some Hersey numbers.

Defining the Stribeck curve of the untextured case to be $s^0(H)$ and the textured Stribeck curve of dimple radius or depth $D_{r/d}$ to be $s^{D_{r/d}}(H)$, we can calculate the total change in the coefficient of friction over the whole Hersey number spectrum as a function of dimple radius as

$$M_f(D_{r/d}) = \int_{\forall H} \left( s^{D_{r/d}(H)} - s^0(H) \right) dH$$

We report the values of $M_f(D_r)$ in figure S3 for multiple values of dimple depth of the fully dimpled case compared to the untextured case. We report also the values of $M_f(D_d)$ in figure S4 for multiple values of dimple radius of the fully dimpled case compared to the untextured case.

It is clear from figures S3 and S4 that the overall change in COF is an order of magnitude greater by changing dimple radius, when compared to the variation observed by changing the dimple depth. The larger effect of the dimple radius was further confirmed by looking at the position of the Stribeck curve minimum as a function dimple radius and depth, when compared to the untextured case, as it is reported figure S5. Therefore, it is possible to conclude that the dimple radius is a better parameter for optimization, owing to its effects in the Stribeck curve.

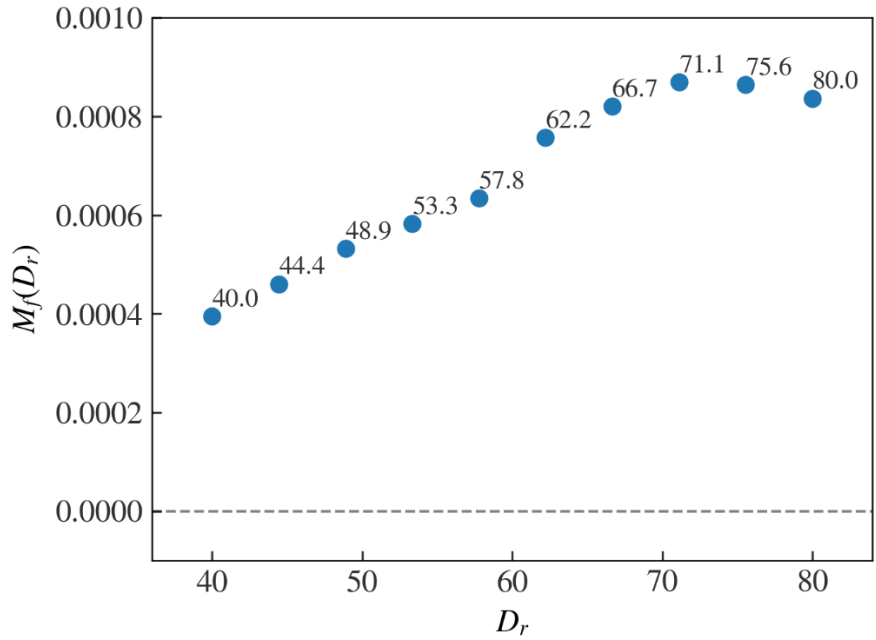

*Figure S3* Total coefficient of friction difference between dimpled cases with radius $D_r$ and the untextured case defined for the mixed and hydrodynamic regions.

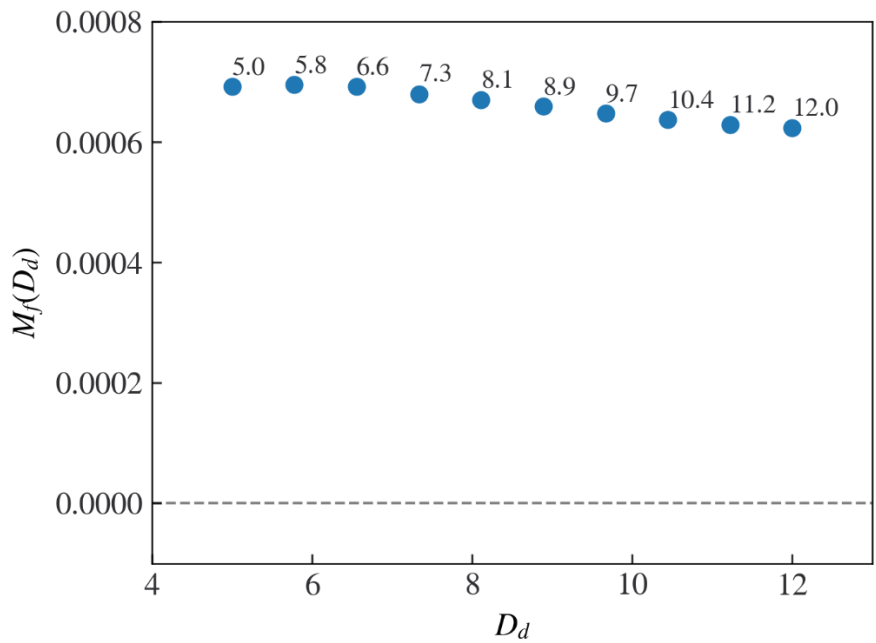

*Figure S4* Total coefficient of friction difference between dimpled cases with depth $D_d$ and the untextured case defined for the mixed and hydrodynamic regions.

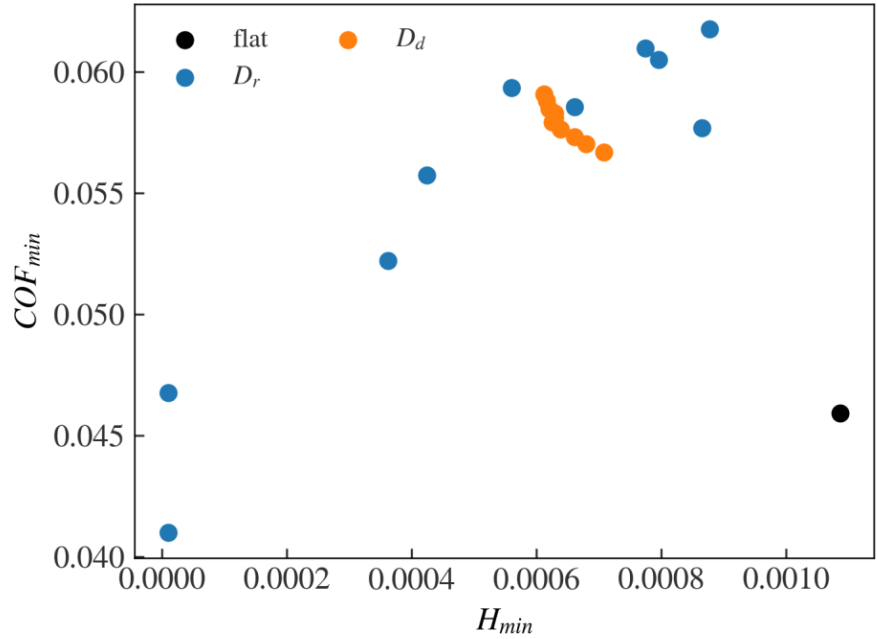

*Figure S5* Minimum position for the untextured case in black, for the radii cases in blue and for the depth cases in orange. A difference in impact is apparent from the relative movement of the minimum in terms of the relevant parameter.

## Section 3 – Parameters used in data generation for the deep neural network training set

*Table 1* Set of parameters used during the generation of training data for the development of the deep neural network used to solve both the forward and inverse problems of texture optimization.

| | | |
|---|---|---|
| $n_x$ | Number of elements in x direction | 100 |
| $n_y$ | Number of elements in y direction | 100 |
| $L_x$ | Physical dimension of the texture in the x direction | 1000 μm |
| $L_y$ | Physical dimension of the texture in the x direction | 1000 μm |
| $h_{min}$ | Initial minimum separation between the two contacting surfaces | 0.25 μm |
| $E_0$ | Parabolical edge height | $2 \times 10^{-5} \mu m^{-1}$ |
| μ | Lubricant viscosity | $0.035\ Pa \cdot s$ |
| F | External load applied to the surface contact | 1.5 N |
| $\eta k\sigma$ | Surface roughness parameters | 0.00128 |
| $E$ | Young modulus | $1.293 \times 10^6 \ N/m^2$ |
| $\sigma_E$ | Eyring sheer stress | $1.974 \times 10^1 \ N/m^2$ |
| $D_d^0$ | Dimple depth used in data generation | 6 μm |
| $D_r^{set}$ | Dimple radius set used in data generation | {40,44,48,52,56,60} μm |
| $H^{range}$ | Hersey number range used in data generation | $[10^{-5}, 10^{-2}]$ |
| $(D_x, D_y)$ | Dimensions of the dimple map | (5,5) |